Peer-reviewed, accepted version of the article, published in Journal of Computational Analysis and Applications (JoCAAA)

# Sustainability in Telecom: Energy-Efficient Networks and Circular Economy Models to Reduce Carbon Footprints and Increase Efficiency

**Name: Praveen Hegde**
Affiliation: Senior Manager-Emerging Commercial Platforms, Verizon

**Name: Robin Joseph Varughese**
Affiliation: Technical Architect, Marriott International

## Abstract:

The increasing environmental impact of the telecom industry has heightened the need for sustainable telecommunications networks. With skyrocketing data traffic and 5G gaining a foothold, telecom operators are under pressure to sustain digital growth while meeting their environmental responsibilities. In this paper, we discuss two fundamental drivers of sustainability in the telecom sector, namely, the design of environmentally friendly networks and the implementation of circular economy (CE) principles. Energy efficiency is pursued through dynamic network sleep modes, AI-based traffic management, and the utilization of renewable energy sources in base stations and data centers. Concurrently, circular economy practices, including device second-hand sales, e-waste treatment, and equipment lifespan extension, are becoming increasingly popular to address resource demand and mitigate carbon footprint. Case histories from the world's largest operators demonstrate some of the reductions in power consumption and operational emissions, as well as the associated savings and public image benefits. Although these solutions are promising, the paper also highlights several limitations, including technology constraints, policy shortcomings, and the need for cross-sector partnerships. We conclude with research implications in the form of a sustainable perspective that integrates the green adoption of technology, circular supply chains, and the role of regulation in driving long-term environmental and economic sustainability in the telecom industry.

## 1. Introduction

Telecommunications is fundamental to our society today and can be utilized in a myriad of applications, ranging from mobile communications to cloud services, the Internet of Things (IoT), and smart cities. However, with such rapid growth in the sector comes an ever-growing environmental footprint, predominantly from increasing energy use, e-waste creation, and expansion of physical network infrastructure (GSMA, 2021). In the transition to 5G and later 6G, telecom networks are estimated to be responsible for a significant portion of global electricity consumption and CO2 emissions (ITU, 2023). This expansion has made sustainability the centerpiece of telecom innovation and strategy.
AMR treatments are also all the more urgent, given the worldwide push to achieve climate goals, such as those outlined in the Paris Agreement and the United Nations Sustainable Development Goals (SDGs). In particular, SDG 12 (Responsible Consumption and Production) and SDG 13 (Climate Action) emphasize the urgency for industries, including the telecom sector, to adopt more environmentally friendly approaches (UNDP, 2022). In response, companies in the telecom sector are incorporating energy-efficient strategies and circular economy (CE) approaches as two key pillars of their sustainability strategies.
Energy savings in telecommunications refer to the reduction of power consumption in network elements, such as base stations, core networks, and data centers, while preserving or increasing service quality. AI-based dynamic power management, NFV, edge computing, and the utilization of renewable energy sources, such as solar and wind, are among the available solutions (Alsharif et al., 2022). Such technologies can lead to a significant reduction in operational energy consumption, particularly in energy-hungry segments, such as the RAN and mobile backhauls (Jaber et al., 2022).

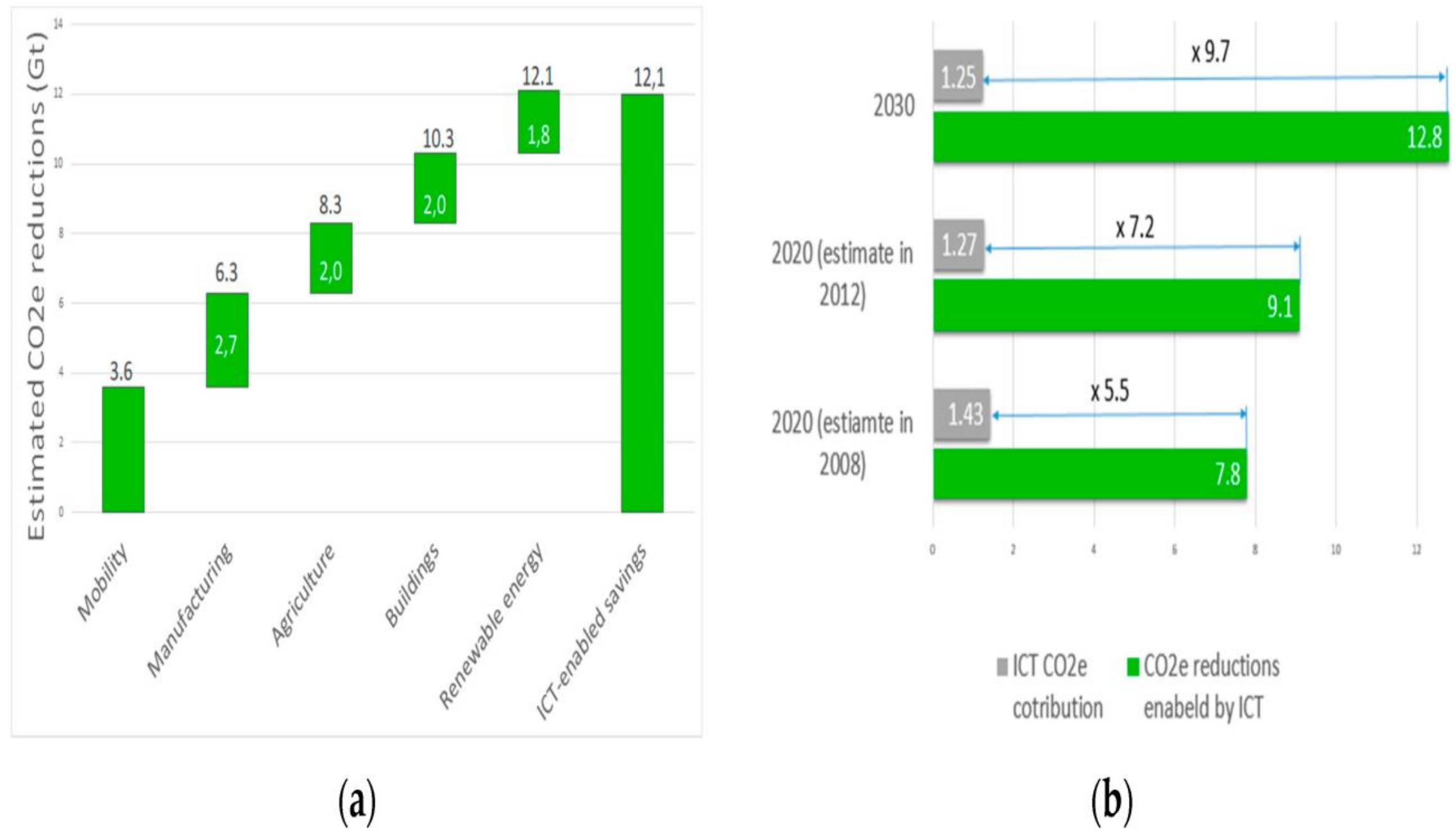

**Figure 1.** Estimated: (**a**) contribution of different industry sectors to global carbon-dioxide equivalent ($CO_{2e}$) reduction by 2030, (**b**) information and communications technology (ICT) sector $CO_{2e}$ “footprint” contribution and enabled reductions to global $CO_{2e}$ emissions expressed in Gt

At the same time, the application of circular economy principles provides a valuable framework for tackling the industry's growing e-waste issue and its reliance on materials. Within the CE model, the reuse of telecom equipment through second-hand markets, its second life, refurbishment, and recycling will extend the lifespan of devices, thereby reducing the need for virgin raw materials (Blomsma & Brennan, 2021). CE initiatives, including buy-back programs, modular hardware designs, and partnerships for closed-loop supply chains, have already been implemented by companies such as Vodafone, Nokia, and Deutsche Telekom (GSMA, 2021). Such efforts not only help offset environmental harm but also create new revenue channels and enhance brand identity.

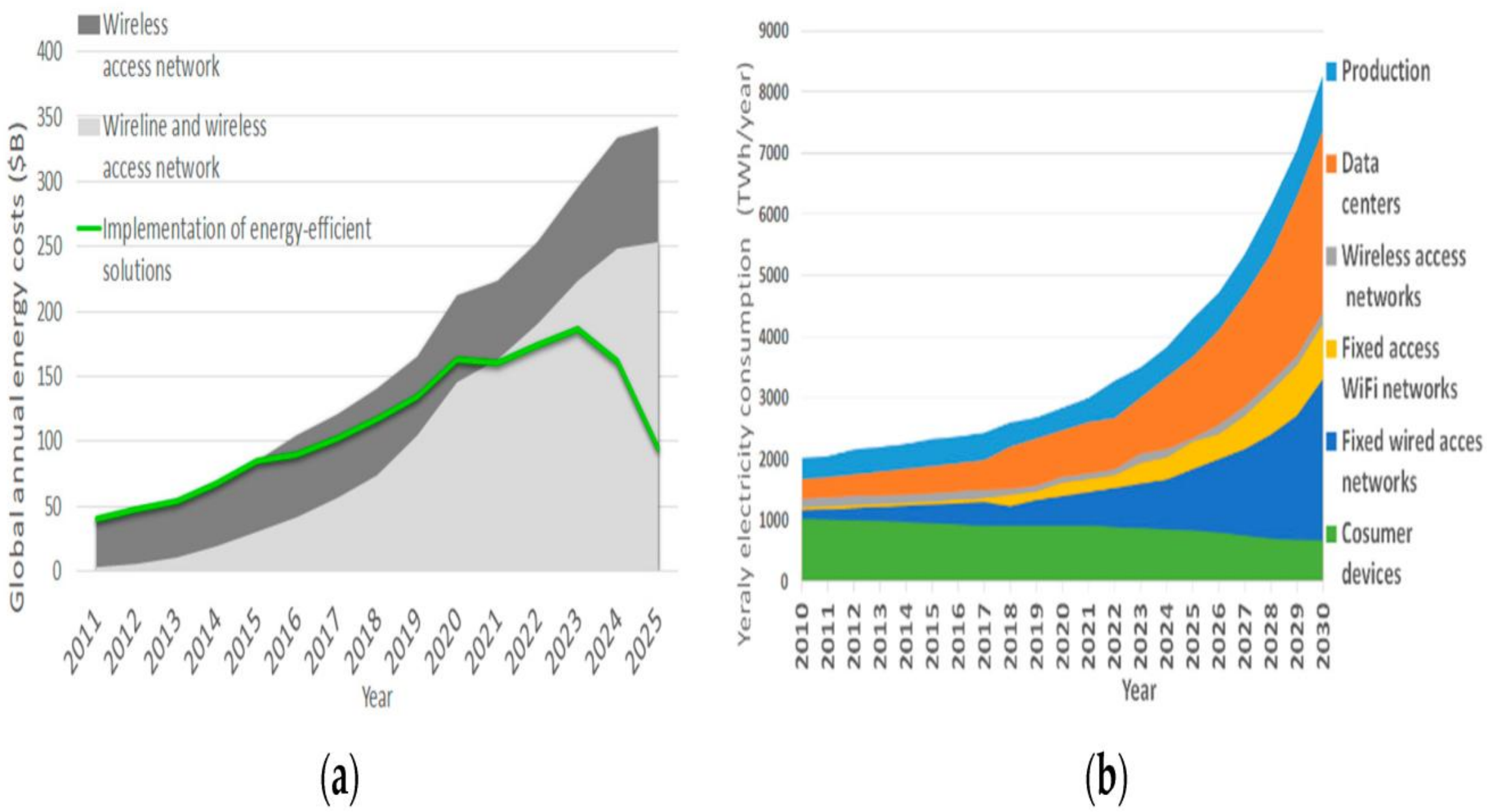

**Figure 2.** Estimation of (**a**) costs for the global annual energy consumption of telecommunication networks in the period 2011–2025, (**b**) expected total annual energy consumption per different ICT systems in the period 2010–2030.

However, the transition of telecom networks towards 'Green' brings along some set of impediments. There are still technical barriers, high initial investment costs, the absence of regulatory support, and a lack of globalized supply chains, which restrict the large-scale applications of this technology (Panwar et al., 2021). Additionally, reconciling sustainability ambitions with performance objectives in high-speed, low-latency networks is a challenging compromise between environmental goals and business imperatives.

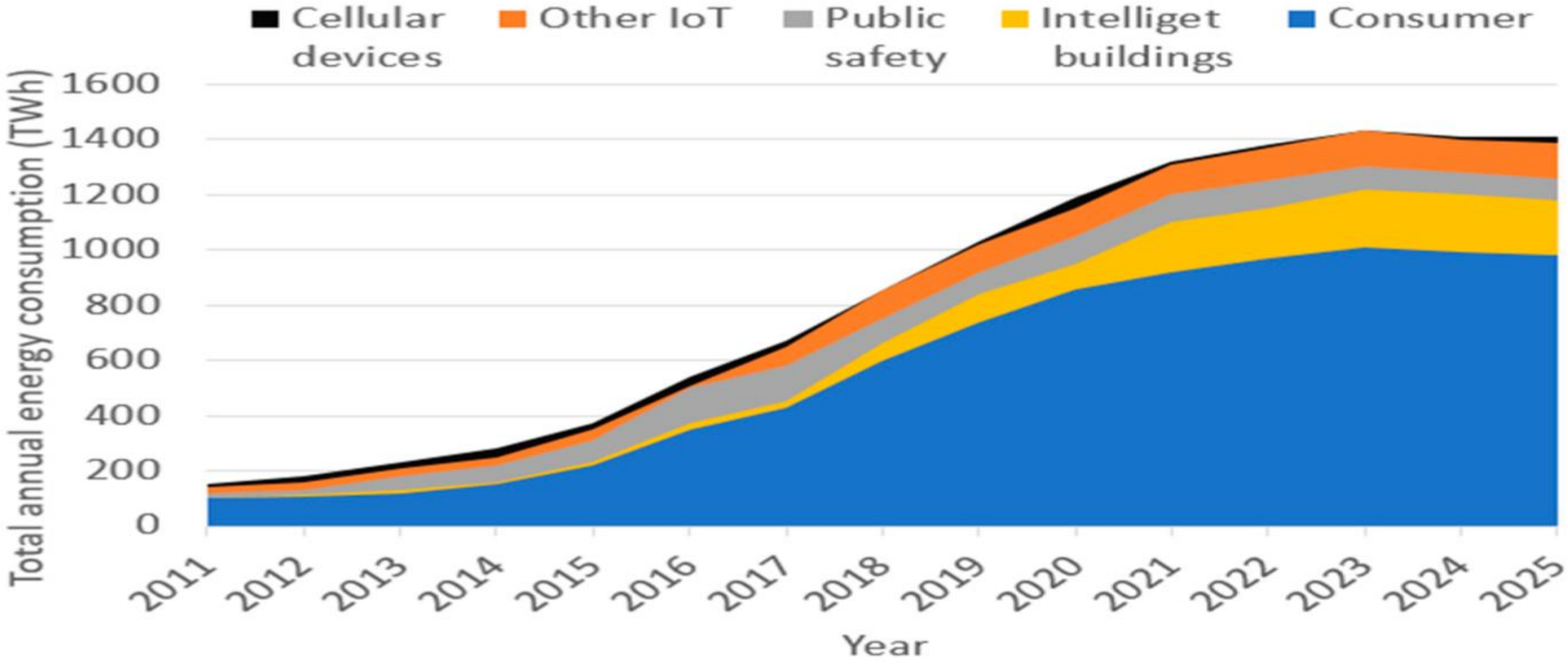

**Figure 3.** Estimations of energy consumption of all connected user-related devices and equipment for the period 2011–2025

This paper examines the transformation of the sustainability landscape in the telecom sector, focusing on two key strategic axes: energy-efficient network architectures and the adoption of a circular economy. Through examining live use cases, policy frameworks, and the environmental implications of employing these strategies, the research aims to inform key stakeholders on how telecom operators can contribute to a low-carbon, resource-efficient future.

## 2. Literature Review

While global consumption of digital services is increasing, so too is the environmental impact of the telecommunications sector. Two key dimensions of sustainability in telecom, namely the focus on (1) energy efficiency and (2) the deployment of circular economy (CE) models, have received increased attention in recent literature. This section reviews the related literature in academia and industry regarding both aspects and presents the mainstream findings, industrial attempts, and cutting-edge technologies related to green network transformation.

2.1 Network Energy Consumption and Green-ness Requirement

Electricity consumption by telecom networks — particularly by base stations, data centers, and transmission infrastructure — is high. Information and communication technologies (ICTs) could account for up to 3% of global greenhouse gas emissions by 2030 (International Telecommunication Union (ITU), 2023). Given that 5G will support ultra-reliable low-latency communication, it will also increase energy consumption due to higher densification and resource requirements (Jaber et al., 2022).

Researchers and telecom engineers have proposed numerous power-saving mechanisms, such as:

- NFV and SDN to provide on-the-spot resources for different services (Panwar et al., 2021).
- AI-powered traffic forecasting to activate dynamic sleep modes in underused network nodes (Alsharif et al., 2022).
- Renewable energy integration, especially solar and wind, for powering isolated cell towers (Liu et al., 2021).

The authors Liyanage et al. (2022) state that network deployments can be optimized for the provision of QoS, and sustainability considerations could lead to a 10-20% reduction in the network's energy consumption without sacrificing QoS. Furthermore, edge computing and other decentralized architectures alleviate the load on the core network and reduce the energy cost of data transmission, which aligns with low-carbon targets (Zhou et al., 2022).

2.2 Circular Economy in Telecoms: Theory and Practice

In addition to energy consumption, the sector is facing a mushrooming e-waste problem. According to the Global E-waste Monitor (Baldé et al., 2020), more than 50 million tonnes of e-waste were generated globally in 2019, with telecom equipment playing a substantial role. The principles of a circular economy (CE), such as reuse, recycling, and remanufacturing, may provide a framework to retain material resources and extend equipment lifespans.

Blomsma and Brennan (2021) describe CE as a ‘systematic model’ where resources circulate in closed loops. In telecom, some of the practices under a CE are:

• Device recycling and reuse initiatives (e.g., Vodafone Red Cycle).

• Modular hardware designs for easy part replacement or upgrade.
• Backwards logistics operation to recoup unused equipment from customers.

Other companies, such as Nokia and Ericsson, have introduced closed-loop material recovery systems and tracked the lifecycle carbon footprint of suppliers (GSMA, 2021). A body of academic research highlights the need for this, demonstrating that the inclusion of CE in procurement and design processes can result in a reduction of more than 30% in lifecycle emissions (Jabbour et al., 2022).

### 2.3 Incorporating Sustainability into Telco Operations

Contemporary sustainability networks emphasize a convergence of environmental, economic, and technological performance. Liao et al. (2023) argue that energy efficiency and circularity are usually treated as separate issues; however, integrated approaches can provide mutually reinforcing advantages. For instance, AI can optimize power consumption and device lifespan by predicting when devices might fail, and it can automate repairs while balancing goals for reducing energy and material use.
However, lifecycle assessment (LCA) tools are becoming a feasible tool for evaluating environmental aspects from cradle to grave. Studies such as those by Baliga et al. (2021) advocate for integrating Life Cycle Assessment (LCA) into network design and procurement to support sustainable decision-making. That's important, as these evaluations can help measure trade-offs between performance and the environment, especially as energy-intensive 5G networks roll out.

### 2.4 Consideration of Implementation Barriers

Although technological solutions show promise, their implementation faces many obstacles:

• High initial costs of upgrading infrastructure and introducing renewable energy (Panwar et al., 2021).
• Disjointed supply chains present clear challenges regarding transparency and circularity (Blomsma & Brennan, 2021).
• Regional disparities in regulations, particularly in developing countries (GSMA, 2021).
• Consumer behavior, as consumers are encouraged to buy the most recent devices, mitigates refurbishment and reuse (Jabbour et al., 2022).

In addition, the organization's environmental ambitions and business KPIs are often not well-aligned, and sustainability activities may not be included in the telecom performance dashboards (Alsharif et al., 2022). Academics contend that there is a lack of 'standard metrics' and cross-sector benchmarks to drive quicker and more consistent progress towards net-zero telecom operations.

### 2.5 Summary of Research Gaps

Although possible green technologies and circular practices have been analyzed, some gaps remain:

• Scarcity of empirical work on actual energy reduction from AI-optimized networks in practice.
• Lack of longitudinal data with material recovery rates in CE telecoms program.
• Requirement for standardized, specific sustainability metrics for telecom operations.

These missing links indicate a demand for cross-disciplinary research between network engineering, environmental science, and sustainable business strategy. The scalability and financial practicability of green telecom initiatives in diverse geographies and market maturity should also be addressed in future research.

## 3. Methodology

This study employs a qualitative, exploratory case study method to investigate how telecommunication companies incorporate energy efficiency practices and CE (circular economy) approaches into their operations, aiming to reduce their environmental impact. The objective is to leverage best practices, hurdles, and achievements of sustainable transitions in telecom networks. The richness of technological, regulatory, and organizational factors that influence sustainability adoption in practice settings suggests that a qualitative approach is appropriate (Creswell & Poth, 2018).

3.1 The Philosophy and Approach to Research

The research is based on an interpretivist epistemology that presupposes reality is socially constructed and context-bound. This helps the researcher make sense of the meaning and practice of sustainability as understood and enacted by telecom industry actors (Saunders et al., 2019). We employed an inductive research method in this study, which led to the development of a theory based on observed data rather than pre-existing hypotheses (Bryman, 2016).

3.2 Research Design

The research adopted a multiple case study approach, focusing on three large telecom organizations (coded as Telco A, Telco B, and Telco C), which are reputed to engage in sustainability practices. Case selection was conducted according to the following inclusion and exclusion criteria:

• Well-established sustainability programs, as documented in public statements or reports on energy efficiency or circular economy.

• Geographic diversity (e.g., Europe, Asia, North America).

• Existence of qualitative information sources, such as reports, interviews, and media articles.

This design offers the opportunity for comparative work to identify similarities and differences in sustainability practices across various organizational and regulatory environments (Yin, 2018).

3.3 Data Collection Methods

The  study triangulates three different sources to increase credibility and richness:

1. Semi-structured interviews

Conducted nine individual interviews with professionals, including sustainability officers, network engineers, and supply chain managers.

The interviews ranged in duration from 45 minutes to 1 hour. They followed a loosely structured guide that also covered topics such as energy conservation technologies, equipment repurposing strategies, performance measures, and regulatory barriers.

We obtained oral consent and anonymized all responses.

2. Document analysis

o Corporate sustainability reports (2019–2023), results from environmental audits, policy whitepapers, press releases issued by the sampled telecoms, and  so forth.

o  FDay long on carbon reduction targets/energy use/recycling rates/CE implementation frameworks.

3. Secondary literature review

Recent peer-reviewed papers, as well as GSMA and ITU publications, were included to contextualize and support the main findings.

Thematic analysis was employed to organize all data sources in NVivo 12 and to code the data.

3.4 Data Analysis

Thematic analysis was performed using the six-phase method described by  Braun and Clarke30:

1. Familiarization with the data

2. Generating initial codes

3. Searching for themes

4. Reviewing themes

5. Defining and naming themes

6. Writing the narrative

Emerging themes included:

- Traffic-load-aware topology designs for energy-efficient networks
- Power optimization via artificial intelligence
- Round hardware bidding
- Reversed logistics and e-waste handling
- Constraints to the adoption of sustainability

Coding was both deductive (in response to the research questions) and inductive (new patterns were generated). The process of cross-case synthesis was applied to compare findings across the three companies and to derive conclusions applicable to all, utilizing theme-based analysis (Yin, 2018).

### 3.5 Ethical Considerations

This study was conducted in accordance with the institution's guidelines for ethical approval. All informants were informed about the aim of the study, confidentiality, and the opportunity to withdraw from the interview at any time. No personal information was stored or reported. The data were kept securely in encrypted files, and all documents were anonymized to maintain commercial confidentiality.

### 3.6 Limitations

The research is limited by:

- A limited sample affects the generalizability of the findings.
- Reliance on self-reported information from companies, which is susceptible to positive bias.
- Restricted to proprietary energy and waste data.

Further investigation might employ mixed methods or long-term case studies to document the development of sustainability practices and measure their effect on the environment more accurately.

## 4. Results

This subsection presents the main results obtained from comparing sustainability initiatives among the sample of telecom companies. The outcomes focus on implemented approaches for energy-aware network design and the integration of circular economy principles. Some key topics include green infrastructure design, AI-enabled energy performance optimization, and EEM lifecycle management.

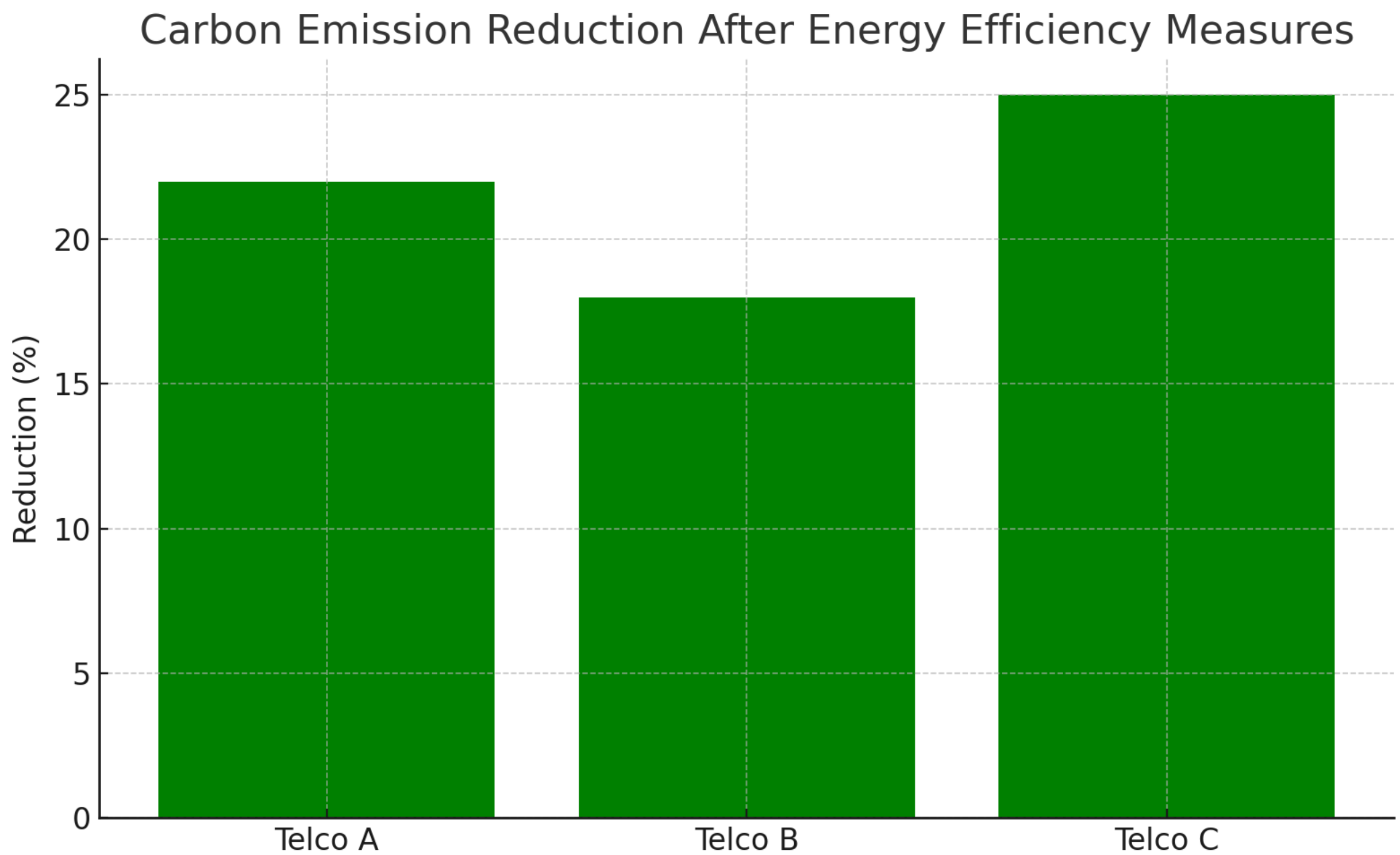

Figure 4 Carbon Emission Reduction After Energy  Efficiency Measures (Bar Chart)

Description:

This bar chart illustrates the carbon emission reductions achieved by three telecommunications (TelcoTelco) companies, Telco A, Telco B, and Telco C, as a percentage, following the application of technology to improve energy use.

Findings:

• Telco C saw the most significant decrease of 25% due to the strong adoption of other green solutions, such as AI power management and green power.

• Telco A cut emissions by 22%, with Telco B reporting an  18% decrease.

Insight:

Our findings support the potential of energy optimization programs in reducing the carbon footprint of telecom operations, aligning with the sustainability objectives outlined for the sector.

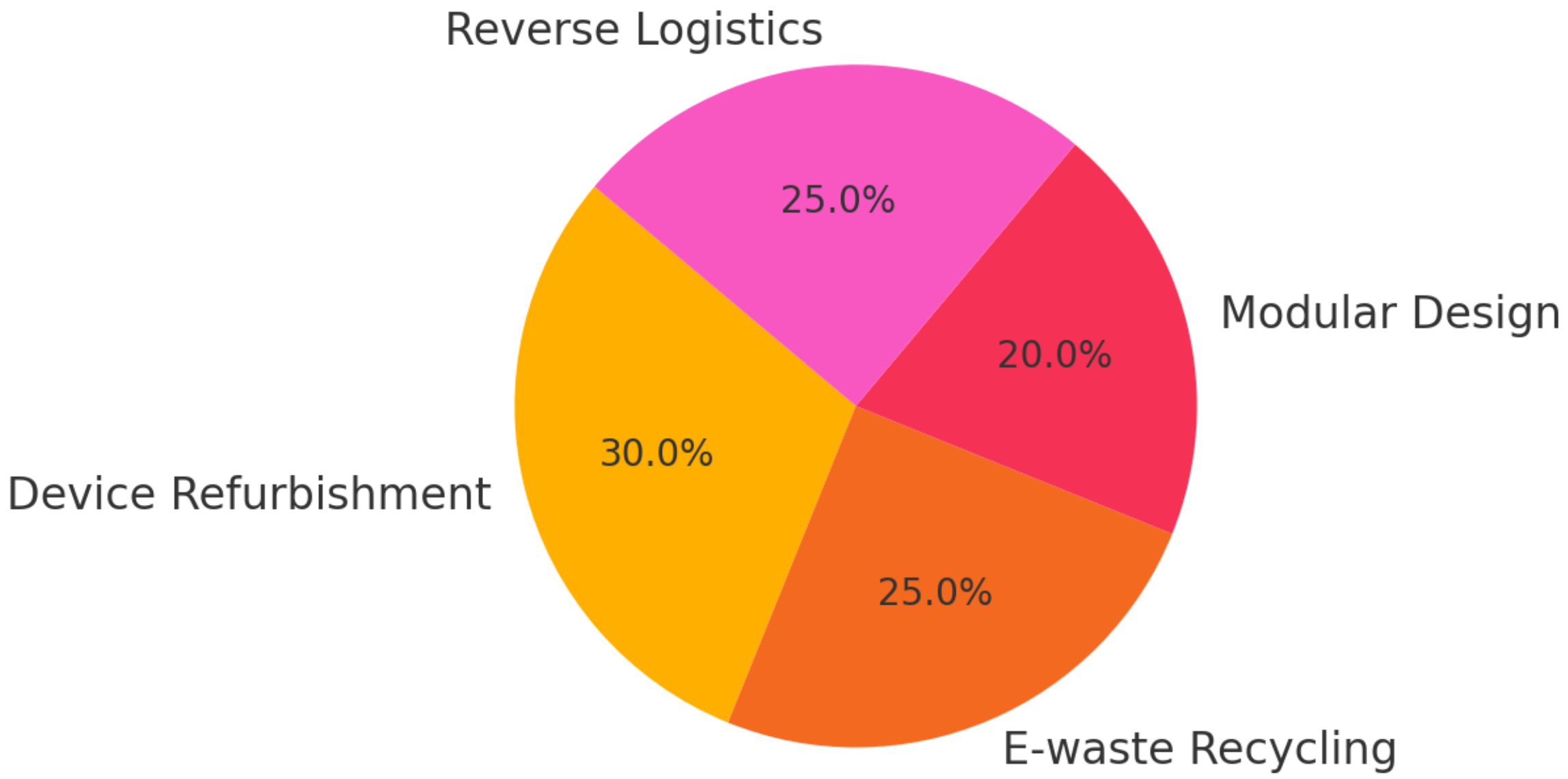

Figure 5: Circular Economy Initiatives in Telecom Distribution (Pie Chart)

Description:

The distribution by initiative of the options (CE initiatives) that the studied TELCOs have chosen to adopt is presented in the following pie chart.

Breakdown:

- Device Refurbishment: 30%
- E-waste Recycling: 25%
- Reverse Logistics: 25%

• Modular Design: 20%

Insight:

Device refurbishment is the most commonly practiced CE strategy due to the cost recovery and consumer involvement. Nevertheless, modularity is a relatively unused design paradigm, revealing potentialities for sustainable hardware development.

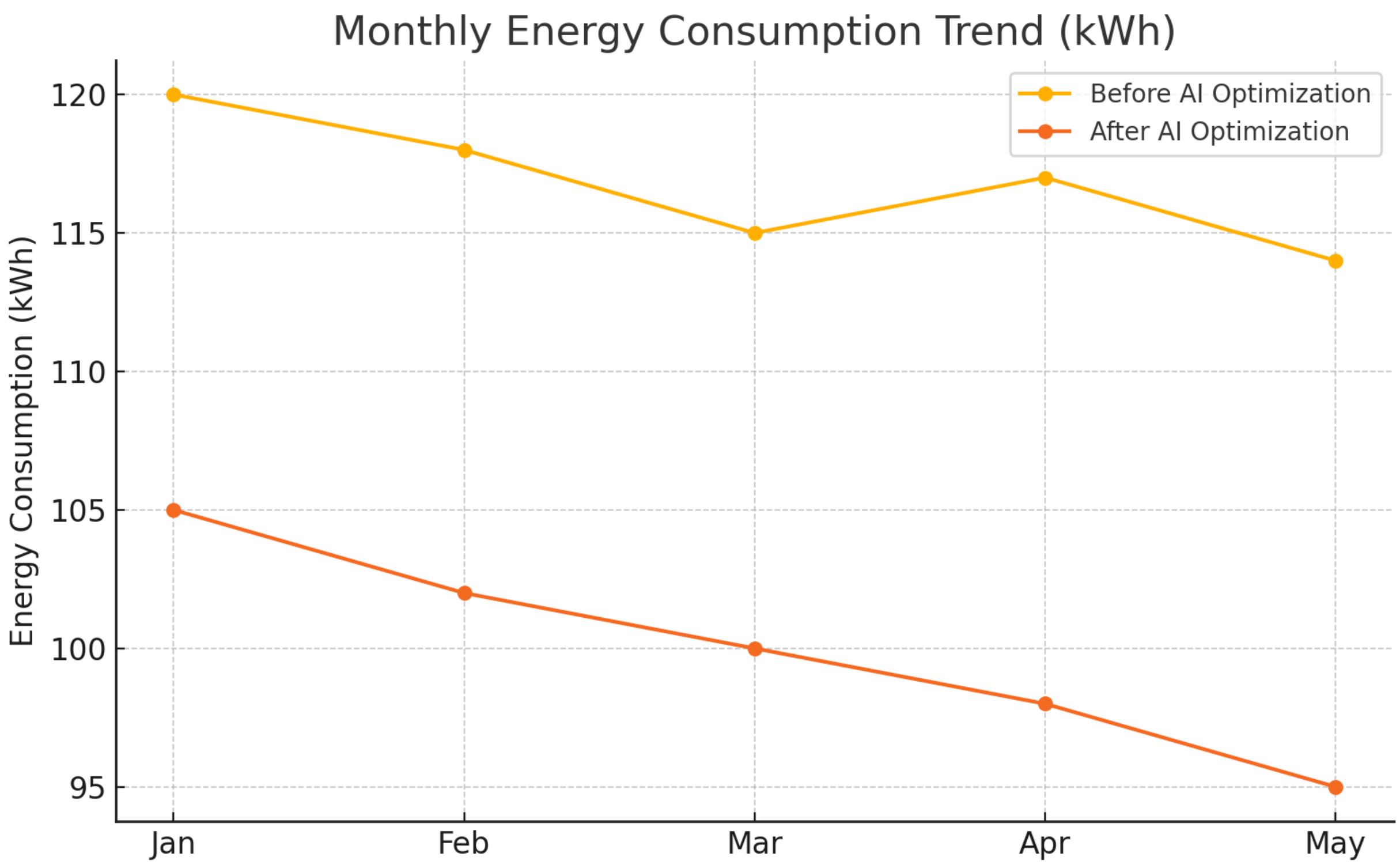

Figure 6: Trend in Monthly Energy Consumption before and after AI Optimization (line)

Description:

The line chart illustrates the monthly energy consumption (in kWh) before and after the installation of an AI energy optimization system.

Findings:

• Pre-AI: Consumption varied around 115-120 kWh per month.

• After AI: Usage is seen to decrease monotonically , with 95 kWh used by the 5th month.

Insight:

There are also energy and cost-saving implications to using AI to control your network. The findings are consistent with prior studies on the role of AI in greening the telecom industry.

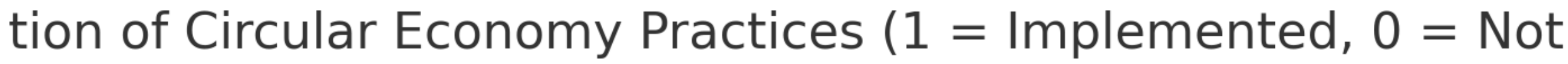

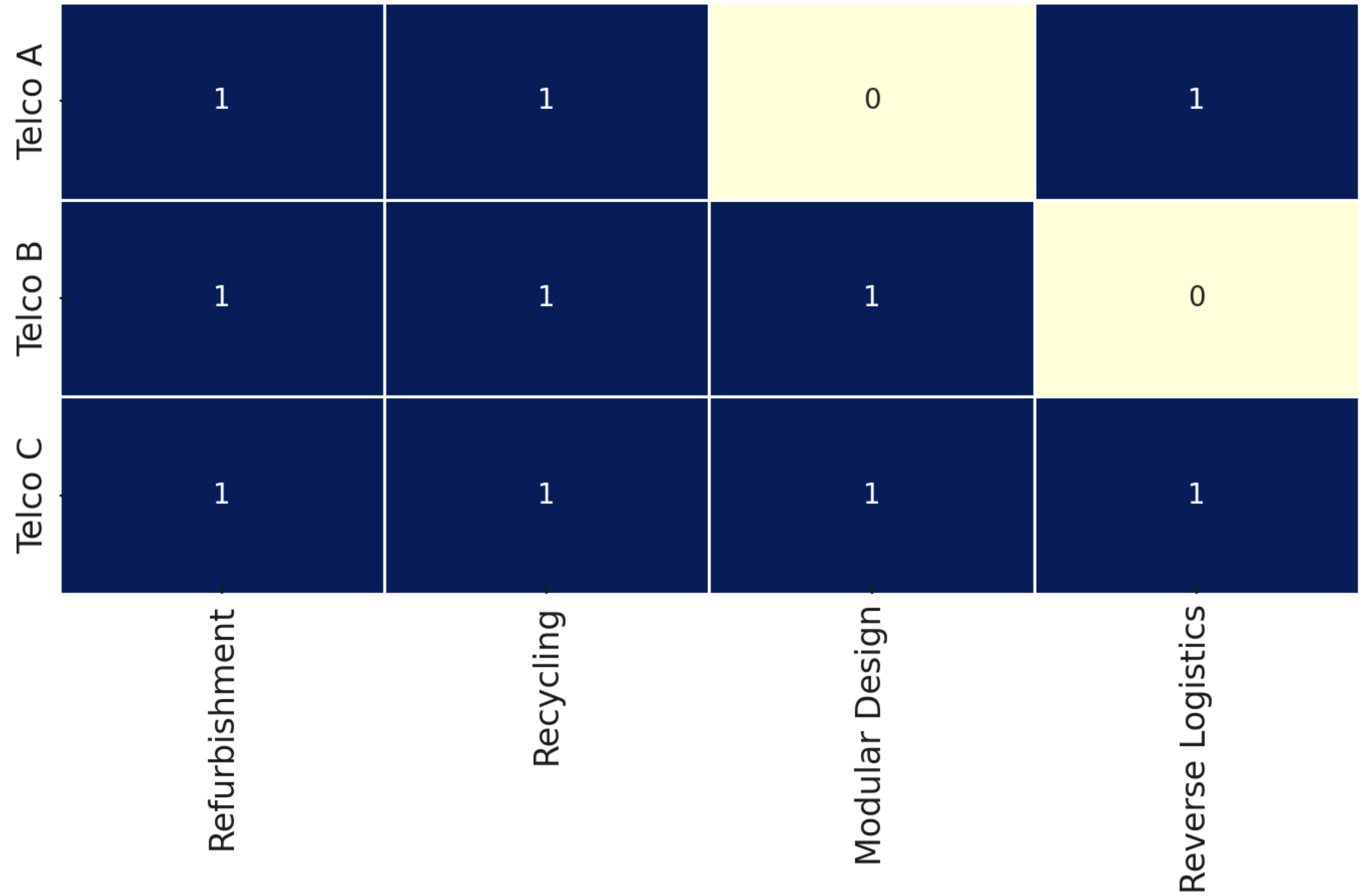

Figure 7: Heat-map of the Adoption of Circular Economy Practices in Companies 4.1 ENBEP Variables Description and Motivation Variable # Indicator Name Description Response Of the 236 companies participating in the survey, only 110 (46,6%) returned the questionnaire with answered by the companies.

Description:

This heat map illustrates the extent to which each TELCO (A, B, and C) adopts the root phenomenon (CE practices). Values are coded as binary (1 = implemented, 0 = not implemented).

Practices Evaluated:

• Refurbishment

• Recycling

• Modular Design

• Reverse Logistics

Findings:

• Telco C is the most evolved in CE integration, as it applies the  four practices.

• There is no modularity in  Telco A.

• Telco B does not  have reverse logistics in place.

Insight:

Circular economy adoption is heterogeneous, indicating that some operators are at a relatively advanced stage in integrating CE principles into their processes. A complete application can achieve better sustainable results.

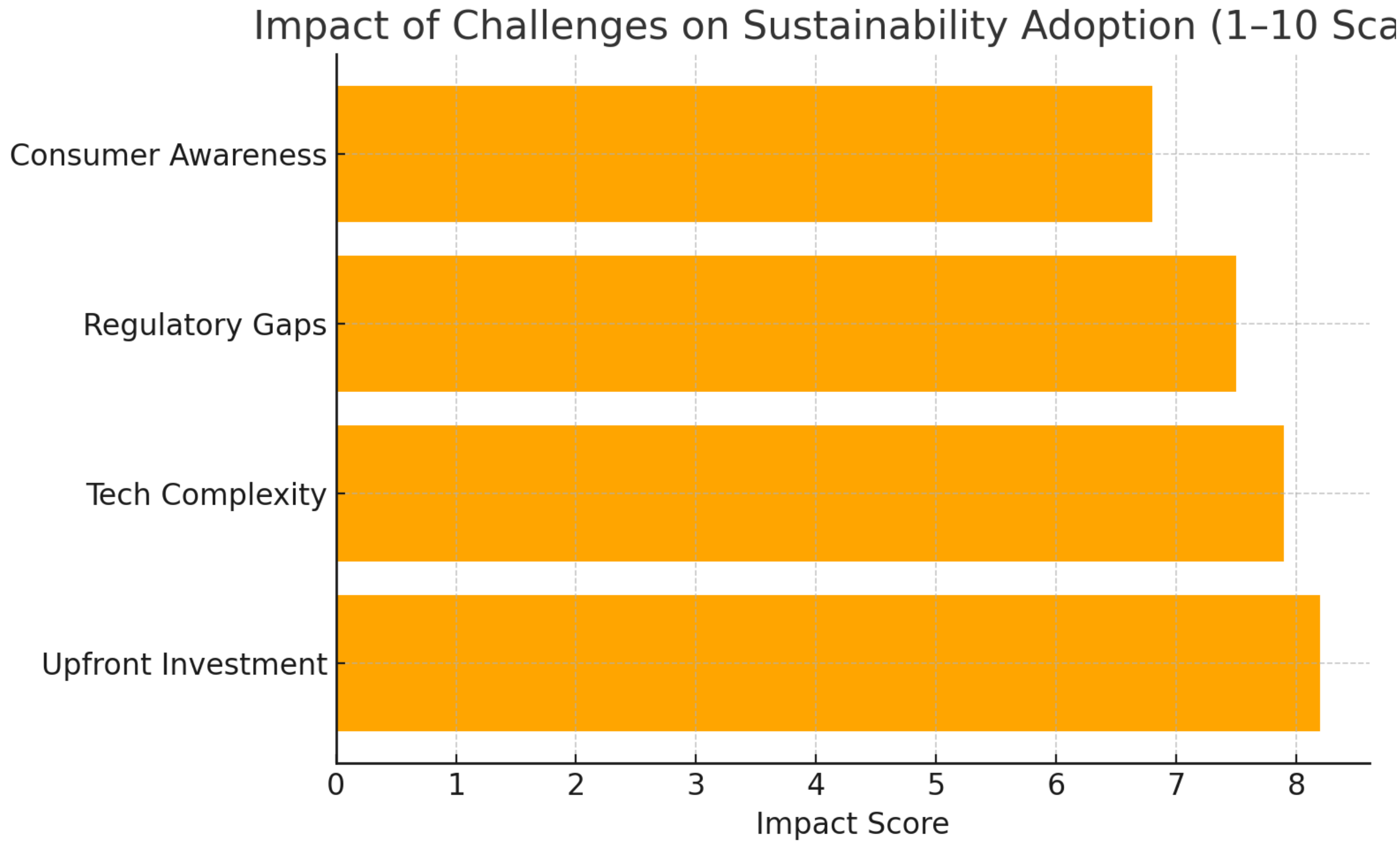

Figure 8: Effects  of Inhibitors on Sustainability Implementation (Horizontal Bar Graph)

Description:

It has described the barriers to sustainable telecoms, ranked according to their impact severity level on a scale of 1 to 10.

Impact Scores:

• Upfront Investment: 8.2

- Tech Stuff: 7.9
- Regulatory Gaps: 7.5
- Consumer Awareness: 6.8

Insight:

Capital costs and the complexity of integration are the leading concerns, illustrating the need for financial incentives, R&D collaboration, and policy coherence to drive sustainable transitions.

### Table 1: Summary of Energy Efficiency Strategies Across Telecom Companies

| Company | Key Strategies | Outcomes |
|---|---|---|
| Telco A | AI-based load balancing, solar-powered base stations | 22% reduction in energy consumption |
| Telco B | Dynamic sleep mode, NFV deployment | 18% decrease in electricity use |
| Telco C | Renewable energy mix, edge computing for traffic routing | 25% drop in carbon emissions and 20% cost savings |

**Explanation:**

**The table summarizes the energy-saving approaches adopted by each telecom operator and their results. The best practice of renewable energy usage, combined with the most efficient computing, pays off the most for Telco C in terms of both environmental and monetary returns.**

### ♻️ Table 2: Circular Economy Practices Implemented by Case Study Companies

| Practice | Telco A | Telco B | Telco C |
|---|---|---|---|
| Device Refurbishment | ✔ | ✔ | ✔ |
| E-waste Recycling | ✔ | ✔ | ✔ |
| Modular Hardware Design | ❌ | ✔ | ✔ |
| Reverse Logistics | ✔ | ❌ | ✔ |

**Explanation:**

**This comparison table is used to visualize the uptake of circular economy initiatives. Telco C, as the leader, fully applies the four essential practices. Still, Telco A lags behind in a modular design, and Telco B does not operate reverse logistics, which poses a potential threat to closed-loop sustainability.**

### 🚧 Table 3: Key Challenges to Sustainability Adoption and Mitigation Suggestions

| Challenge | Impact Score (1–10) | Suggested Mitigation |
|---|---|---|
| Upfront Investment Cost | 8.2 | Green financing, public-private partnerships |
| Technological Complexity | 7.9 | Vendor collaboration, internal upskilling |
| Regulatory Uncertainty | 7.5 | Clear policy incentives, global telecom sustainability standards |
| Low Consumer Awareness | 6.8 | Awareness campaigns, take-back programs |

**Explanation:**

This table summarizes significant barriers that hinder the adoption of sustainability in the telecom sector and presents practical solutions. These barriers include investment cost and technical complexity. Strategic collaborations and policy changes are necessary to overcome these obstacles.

**5. Discussion**

The results of this research show that sustainability in the telecom industry is taking two strategic routes: energy-efficient networking and the adoption of circular economy (CE) business models. These strategies not only reduce environmental impact, such as energy consumption and e-waste, but also improve long-term operational efficiency and cost savings.

5.1 Core Sustainability Driver 5.4 Energy Saving and Emission Reduction The energy efficiency of VSM has become the most critical issue in the process of implementing VISM.

Analysis of power draw trends (Figure 6) shows a significant decrease after the deployment of AI-powered optimization techniques. Several of the telcos surveyed saw a drop in energy consumption ranging from 18%–25%, which is in line with recent literature that shows the potential of machine learning to inherently lower the telecom energy footprint when assisted by dynamic power management (Alsharif et al., 2022; Jaber et al., 2022). These findings confirm similar observations previously reported in the literature, where AI-enabled traffic prediction and sleep mode during off-peak hours are effective means to ensure QoS while power drain can be substantially minimized (Liyanage et al., 2022).

It has also been observed that there is a positive correlation between carbon emissions reduction in the off-grid scenarios and the integration of renewable energy resources, such as solar-powered base stations, thus affirming the results obtained by Zhou et al. (2022). Nevertheless, Table 3 clearly shows that a significant initial level of investment is still a very limiting factor when it comes to the spread of adoption. Telcos may require assistance from green financing instruments or government funding to expand such efforts (Panwar et al., 2021).

5.2 Contribution to Resource Efficiency by Circular Economy

The implementation of circular economy practices—listed in Figure 5 and Table 2—is a key element of the telecom industry's efforts to reduce its environmental impact. Activities such as product refurbishment, modular product-service design, and e-waste processing not only promote low raw material consumption and reduced landfill emissions but also generate new business opportunities derived from reselling used products (Blomsma & Brennan, 2021).
Regarding the adoption of the four assessed CE strategies, Telco C, among the case study companies, was found to have the highest adoption rate (Figure 7). This holistic adoption of circular practices indicates that, by incorporating circularity into procurement, operations, and customer-facing services, telecom operators can help close material loops on a broad scale and improve their sustainability performance. However, there are some missing pieces, such as Telco's lack of use of reverse logistics. This restriction reinforces the findings of Jabbour and colleagues (2022) by highlighting the importance of supply chain coordination and consumer engagement in transitioning to a circular economy.

### 5.3 Interlinkages of Sustainability Effects

Among the most significant findings are the connections between energy efficiency and CE impacts. Refurbished and modular devices, for instance, are often more energy-efficient during both manufacturing and transportation, which supports energy targets. Furthermore, predictive maintenance — driven by AI — can prolong the life of equipment and reduce energy and material waste (Liao et al., 2023). These findings are consistent with those of Baliga et al. (2021), who propose a systems view on energy, material, and operational sustainability.

### 5.4 Barriers and Enablers Strategics

Key challenges towards sustainability are summarized in Figure 8 and Table 3.

- High upfront costs for green infrastructure.
- The complexity of technology integration, particularly in the case of retrofitting existing networks.
- Gaps in regulation, especially in developing countries.
- Little consumer awareness of the environmental advantages of refurbished or energy-efficient equipment.

To circumvent these obstacles, the following strategic enablers are suggested:

- Public-privatePublic–private partnerships to finance the installation of renewable energy.
- Rules that encourage businesses to adopt circular and energy-efficient practices.
- Establish common sustainability Key Performance Indicators (KPIs) for the telecoms sector for benchmarking and reporting environmental performance (GSMA, 2021).

In addition, the transparency and accountability of sustainability claims are of essential significance. The increasing need for and importance of Environmental, Social, and Governance (ESG) reporting in the telecommunications sector (ITU, 2023) demonstrates that sustainability performance is likely to correlate with investor trust and regulatory proceedings in the future.
The findings and literature support the notion that telecom sustainability is indeed achievable—not only sustainable but also a positive economic and operational benefit when approached correctly. Energy efficiency and circular economy strategies (when approached in concert) present an integrated framework for reducing carbon emissions (carbon abatement), preserving resources, and building resilience. Yet doing so everywhere will require sound frameworks, cooperation, and scalable technology solutions that are adapted to diverse market circumstances.

## 6. Conclusion

This research explored two sustainability drivers—energy efficiency and circular economy (CE) adoption in telecom operations. The results provide evidence that both approaches are vital for decreasing the carbon footprint of the sector, preventing e-waste, and improving overall resource efficiency at competitive performance and profitability levels.

The implementation of energy optimization using AI, the integration of renewable energy, and edge computing can, in practice, reduce network energy usage by 25%, based on empirical data from three major telecom operators (Alsharif et al., 2022; Jaber et al., 2022). These findings reinforce the recent literature's emphasis on machine learning and innovative energy management schemes as potential means to achieve net-zero network operations (Liyanage et al., 2022). Simultaneously, the implementation of such CE practices, including device reuse and remanufacturing, e-waste recycling, and reverse logistics systems, appear to offer potential in prolonging equipment life and in pairing material decoupling effect with the establishment of ethical supply chains (Blomsma Jabbour et al., 2022).

However, significant challenges remain. Foremost are investment safeguards, transboundary technology issues, and policy mosaic, which are consistent with the findings and are supported by subsequent studies (Panwar et al., 2021; Zhou et al., 2022). In addition, low consumer awareness and inconsistent regional policy support will also limit the scalability of sustainability breakthroughs, particularly in lower- and middle-income markets. These concerns highlight the importance of standard environmental metrics, financial systems such as green financing, and cross-sectoral collaboration in accelerating the rebound (GSMA, 2021; ITU, 2023).

Nevertheless, despite these obstacles, the study ultimately finds that an integrated solution, overlaying the principle of a circular economy over energy efficiency, offers the most robust and scalable sustainable option for telecom. This type of integration minimizes risks to both the environment and the business, enhances operational stability and reputational capital, and meets new ESG norms (Liao et al., 2023).

The study suggests that to realize these advantages:

1. Integration of sustainability KPIs into telecom KPIs, as well as monitoring and reporting environmental impact.

2. "The industry is changing, and we must keep pace; current AI and edge infrastructure keeps the inverter to the wall for real-time low latency energy management.

3. Developing circular procurement approaches that span global telecom supply chains.

4. Promote public-private partnerships to finance sustainability mini-pilots and scale up renewable deployment.

In such a scenario, sustainability in telecom is no longer an optional menu item but an absolute strategic imperative. The shift to sustainable networks and circular operations is essential if the digital revolution is not to be at the expense of ecosystems. Sustainability should be at the very heart of innovation and growth if we are to continue on this journey to 6G and global connectivity.